\documentclass[conference]{IEEEtran}

\usepackage{amsthm}
\usepackage{times,amsmath,epsfig}
\usepackage[T1]{fontenc}
\usepackage{graphicx} 
\usepackage{color} 
\usepackage{subfigure}
\usepackage{float}
\usepackage{boxedminipage}
\usepackage{multirow}

\usepackage{cite}

\begin{document}

\title{Exploiting Capture Effect in Frameless ALOHA for Massive Wireless Random Access}

\author{\IEEEauthorblockN{\v Cedomir Stefanovi\' c\IEEEauthorrefmark{1}, Miyu Momoda\IEEEauthorrefmark{1}\IEEEauthorrefmark{2}, Petar Popovski\IEEEauthorrefmark{1},}
\IEEEauthorblockA{
\IEEEauthorrefmark{1}Department Electronic Systems, Aalborg University, Aalborg, Denmark\\
Email: \{cs,petarp\}@es.aau.dk\\}
\IEEEauthorblockA{\IEEEauthorrefmark{2}Osaka City University, Osaka, Japan\\
Email: momoda@c.info.eng.osaka-cu.ac.jp\\}
}

\maketitle

\begin{abstract}

The analogies between successive interference cancellation (SIC) in slotted ALOHA framework and iterative belief-propagation erasure-decoding, established recently, enabled the application of the erasure-coding theory and tools to design random access schemes.
This approach leads to throughput substantially higher than the one offered by the traditional slotted ALOHA.
In the simplest setting, SIC progresses when a successful decoding occurs for a single user transmission.
In this paper we consider a more general setting of a channel with capture and explore how such physical model affects the design of the coded random access protocol.
Specifically, we assess the impact of capture effect in Rayleigh fading scenario on the design of SIC-enabled slotted ALOHA schemes. We provide analytical treatment of frameless ALOHA, which is a special case of SIC-enabled ALOHA scheme.
We demonstrate both through analytical and simulation results that the capture effect can be very beneficial in terms of achieved throughput. 
\end{abstract}

\section{Introduction}

The rise of M2M communications introduced necessity for efficient random access mechanisms, motivating new research
approaches that put novel views on the traditional solutions, such as the slotted ALOHA (SA).
One of the promising directions in this respect is the use of successive interference cancellation (SIC) in the slotted ALOHA framework, which enables to exploit collisions and thereby boost the throughput.
The use of
SIC in framed SA\cite{OIN1977} was originally proposed in \cite{CGH2007}.
A systematic treatment of the concept was presented in the seminal paper by Liva~\cite{L2011}, where the analogies between SIC in framed SA and iterative belief-propagation (BP) decoding or erasure-correcting codes were identified.
This opened the possibility to use the theory and tools of codes-on-graphs, laying the foundations of the coded random access.
The ideas of coded random access in a setting with framed SA were further developed in \cite{PLC2011,PLC2011b,LPLC2012,NP2012}, where the main message is
that the use of SIC, coupled with a proper access strategy, grants a throughput that tends to 1 asymptotically i.e., when number of users $N \rightarrow \infty$.

The application of coded random access in the original slotted ALOHA framework~\cite{R1975}, where the users perform access on a slot basis, rather than on a frame basis, was proposed in \cite{SPV2012,SP2013}, introducing the approach of frameless ALOHA.
The operation of frameless ALOHA is inspired by rateless codes \cite{BLMR1998}: the slots are ``added'' to the contention process until the base station decides to terminate the contention; the contention termination criterion can be based, for example, on throughput maximization.
In \cite{SP2013} it was shown that a simple version of the scheme, where the users access the slots with probability that
is uniform both over users and slots, leads to throughput values that are the highest in the reported literature for practical number of users in the range $N \in [50,1000]$.

To the best of our knowledge, the only reference to the capture effect in the literature on coded slotted ALOHA was made in \cite{L2011}, which presents a general modification of the and-or tree evaluation\footnote{And-or tree evaluation is a tool used to assess the asymptotic performance of the iterative BP erasure decoding, i.e., SIC in out setting, see Section~\ref{sec:analysis}.} \cite{LMS1998} that takes into account the capture effect. 
In this paper we make several steps forward.
First, we present a detailed treatment of the capture effect in coded slotted ALOHA, providing a systematic approach for the actual computation of the expressions that constitute the and-or tree evaluation, as compared to \cite{L2011}.
We apply the derived analysis to the Rayleigh fading scenario in frameless ALOHA and show that higher throughput can be achieved in comparison to the simplest communication model that has no capture effect. 
We also perform a simulation-based study and demonstrate that the capture effect can be exploited to improve the throughput in the non-asymptotic case.

The organization of the rest of the text is as follows.
Section~\ref{sec:background} briefly introduces the concepts of the coded slotted ALOHA and capture effect.
Section~\ref{sec:model} provides the system model.
Section~\ref{sec:analysis} presents a detailed derivation of the and-or tree evaluation that takes into account capture effect, and instantiates it for the case of Rayleigh fading and frameless ALOHA.
Section~\ref{sec:results} presents both analytical and simulation-based results, while Section~\ref{sec:conclusions} concludes the paper. 

\section{Background and Related Work}
\label{sec:background}

\subsection{Coded Slotted Aloha}
\label{sec:CSA}

\begin{figure}[t]
 \begin{center}
  \includegraphics[width=0.95\columnwidth]{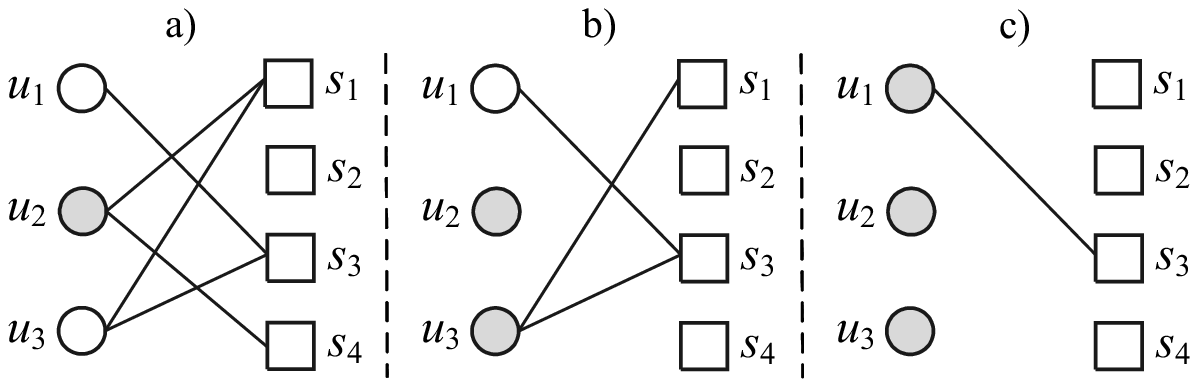}
  \caption{Graph representation of coded slotted ALOHA.}
  \label{fig:graph}
 \end{center}
\end{figure}

A toy example illustrating the principles of coded SA (i.e., SIC-enabled SA) is shown in Fig.~\ref{fig:graph}.
The nodes on the left represent users, the nodes on the right represent slots, and the edges connect users with slots in which their respective transmissions take place.
All transmissions made by a user are replicas of the same packet; we assume that every transmission includes pointers to all other replicas.\footnote{The practical details of sending pointers are out of scope, more information on this topic can be found in \cite{SP2013}.}

In its simplest form, when there is no capture effect, SIC proceeds as follows.
First, the slots containing a single transmission, referred to as \emph{singleton slots}, are identified and corresponding transmissions recovered.
In the depicted example, Fig.~\ref{fig:graph}a), $s_4$ is a singleton slot and the packet of user $u_2$ is recovered from it.
In the next step, using pointers from the recovered transmissions, the slots that contain replicas are identified and the corresponding transmissions removed, i.e., the interference caused by them is canceled. 
This may result in new singleton slots, as depicted in Fig.~\ref{fig:graph}b), where $s_1$ becomes a singleton slot.
The above procedure iterates in the same way, until there are no new singleton slots, or all user transmissions have been recovered.
We note that the described procedure is completely analogous to the iterative BP erasure decoding, where the left nodes represent the original data symbols, the right nodes represent the XOR-encoded symbols, while an edge from a data symbol to an encoded symbol means that the data symbol is XOR-ed within the encoded symbol. 

The SIC algorithm with capture effect is executed in a similar way.
A crucial difference is that, instead operating only on singleton slots, the algorithm also exploits slots where the capture effect takes place. 
Specifically, both in the initial step and in the subsequent iterations, new user transmissions are recovered from: 1) singleton slots, or slots that become singletons due to the interference cancellation, or 2) collision slots, including those that remain collision slots after interference cancellation, in which the capture effect occurs, as explained in the next subsection.
Note that this feature introduced by the capture effect is not present in the analogous code-on-graph with XOR-encoded symbols, as in such analogy the decoding is limited only to singleton slots. 

\subsection{Capture Effect}

When multiple users contend simultaneously for the same slot in SA, it is assumed by default that all transmitted signals in the slot fail to be received due to the collision.
However, the capture effect may take place, that is, the strongest received signal may be successfully received despite the presence of interfering signals from other users in the same slot.
The capture effect in SA related literature has been considered extensively, see\cite{R1975,N1984,GVS1988,WEM1989}.

A typical premise is that a capture can be enabled by having different power levels for different received packets.
These differences are results of propagation phenomena, such as fading, shadowing and the near-far effect.
Stated more formally, the capture effect occurs when the signal-to-interference-plus-noise ratio (SINR) is larger than a predefined threshold, called capture ratio $b$. The actual value of the capture ratio depends on the coding/modulation used in a particular system. 
In \cite{NEW2007,ZZ2012}, it is shown that the capture effect is applicable both in narrowband ($b \geq 1$) and broadband ($0 \leq b < 1$) systems.
The random distribution of the signal powers at the receiver and the SINR criterion to determine the capture probability have been considered in a number of works, like \cite{N1984,ZR1994,ZZ2012}.

The capture effect can be applied both in multi-packet reception (MPR) and SIC enabled receivers.
Typically, the MPR is implemented in broadband systems, where $b<1$, such as e.g. spread-spectrum systems. On the other hand, SIC is used in conventional narrowband systems, where $b\geq1$. In this paper we focus on the latter, narrowband scenario, as an appropriate setting for M2M communications.
We also note that a typical assumption is that the SIC is performed only on the signals received in the same slot, i.e., intra-slot SIC \cite{NEW2007,ZZ2012}.
However, the design features of coded SA allow both for intra- and inter-slot SIC, i.e., the interfering signals are removed from all slots where the repeated transmissions take place, thus ``unlocking'' all the slots where the capture effect can be exploited again.

\section{System Model}
\label{sec:model}

We consider a scenario in which $N$ users contend for the access to the Base Station (BS), where $N$ is assumed known at the BS.
The contention period is divided into $M$ slots of equal-duration, where $M$ is not a priori fixed value.
The BS starts and terminates the contention period by sending the beacon; we assume that the duration of the beacon is one slot.
For each slot, each user decides whether to transmit or not randomly, with a predefined probability termed slot access probability $p_a$. The value of $p_a$ is broadcast by the BS through the beacon that is used to initiate contention and is set to:
\begin{align}
\label{eq:pa}
p_a = \frac{\beta}{N},
\end{align}
where $\beta$ is a suitably chosen parameter, subject to optimization. 
Note that $p_a$ is the same for all users and constant for all slots. Despite its simplicity, this approach yields exceptionally good results, as verified in \cite{SPV2012,SP2013}.

Denote by $u_i$ the $i$-th user, $1 \leq i \leq N$, and by $s_j$ the $j$-th slot of the contention period, $1 \leq j \leq M$.
Further, denote by $|u_i|$ the number of transmission that $u_i$ performs during the contention period and by $|s_j|$ the number of colliding transmissions in slot $s_j$.
Henceforth, we will refer to $|u_i|$ and $|s_j|$ as user and slot degrees, respectively.
From \eqref{eq:pa}, it is straightforward to show that $E[|u_i|]= \frac{M}{N}\beta = ( 1 + \epsilon)\beta $, $1 \leq i \leq N$, where $\epsilon = \frac{M}{N} - 1$, and $E[|s_j|]=\beta$, $1 \leq j \leq M$.
Also, it could be shown that the actual values of $|s_i|$ of $|u_j|$ are binomially distributed, which can be approximated by Poisson distributions for the ranges of $N$, $M$ and $\beta$ of interest:
\begin{align}
\label{eq:lambda}
\mathrm{Pr}[|u_i|=k] = & \Lambda_k \approx \frac{(1+\epsilon)^k\beta^k}{k!} e^{-(1 + \epsilon)\beta}, \; 1 \leq i \leq N, \\
\label{eq:omega}
\mathrm{Pr}[|s_j|=k] & = \Omega_k \approx \frac{\beta^k}{k!} e^{-\beta}, \quad 1 \leq j \leq M ,  
\end{align}
where $ k \geq 0$.

During the contention period, $u_i$ transmits a signal $U_i$.
The BS receives composite signal $S_j$ in slot $s_j$:
\begin{align}
\label{eq:composite}
S_j = \sum_{i=1}^{N} a(i,j) h_i U_i + Z_j, \;  1 \leq j \leq M,
\end{align}
where: 1) $a(i,j)=1$ if user $i$ transmits in slot $j$, is 0 otherwise, and $\mathrm{Pr}\left[a(i,j)=1 \right] = p_a$, $1 \leq i \leq N$, $1 \leq j \leq M$; 2) $h_i$ is the channel coefficient of $u_i$; and (3) $Z_j$ is the noise.
The received powers $P_i = |h_i|^2 E[|U_i|^2]$, $1 \leq i \leq N$, are assumed to be independent and identically distributed (IID) random variables, that depend on the transmit powers, the statistical distribution of the distance between the user and the BS and the stochastic phenomena on the wireless link \cite{ZZ2012}; also, their values do not change during the contention period.

The BS stores all received slots (i.e., the received composite signals) and after each received slot, performs SIC until there are no new degree one slots, or higher degree slots that are exploitable due to the capture effect, as explained in Section~\ref{sec:CSA}.
The above process is repeated until the BS terminates the contention period by sending a new beacon; this effectively and a posteriori determines the value of $M$.
For the sake of simplicity we assume a perfect SIC, i.e., there is no residual interference power remaining after SIC is performed.\footnote{We note that the presented analysis can be extended to include a proportional model of the residual interference power, as shown in \cite{ZZ2012}.}

Finally, we introduce the criterion for contention termination.
Denote by:
\begin{align}
\label{eq:Fr}
F_R & = \frac{N_R}{N},\\
\label{eq:T}
T_I & = \frac{N_R}{M+1}, 
\end {align}
instantaneous fraction of resolved users and instantaneous throughput respectively, where $N_R$ is number of resolved users (the term +1 in the denominator of \eqref{eq:T} takes into account the slot used for the beacon transmission). 
The termination criterion consists of two conditions: the contention is terminated \emph{either} when $F_R \geq V$, \emph{or} when $T_I \geq S$, where $V$ and $S$ are the respective thresholds, chosen such that the expected throughput $\bar{T}$ is maximized.\footnote{A comprehensive treatment of the design of the contention termination criterion can be found in \cite{SP2013}.}

\subsection*{Capture Effect in Rayleigh Fading Scenario}

Here we briefly treat the case when packets arrive through a frequency non-selective Rayleigh fading channel, a scenario for which the results are presented Section~\ref{sec:results}.
We define $X_i$ as a random variable that represents received signal-to-noise ratio (SNR) of user $u_i$, i.e., $X_i = \frac{P_i}{|Z_i|^2}$, and assume that
$X_i$, $1 \leq i \leq N$, are independent and identically exponentially distributed with mean $\bar{\gamma}$:
\begin{align}
p_{X_i} (x) = \left\{ 
  \begin{array}{c l}
    \frac{1}{\bar{\gamma}} e^{- \frac{1}{\bar{\gamma}}x}, & \quad x \geq 0\\
    0, & \quad \text{otherwise}
  \end{array} \right.
\end{align}

A user transmission is captured in slot of degree $n$ when its SINR is larger than a capture ratio $b$, i.e., when:
\begin{align}
\label{eq:cond}
\frac{X}{1+\sum_{j = 1 }^{n-1} X_{i_j}} \geq b,
\end{align}
where $X$ represents the user's SNR, $X_{i_j}$, $1 \leq j \leq n - 1$, are SNRs of the $n-1$ interfering users, and $b \geq 1$.
The condition~\eqref{eq:cond} can be rewritten as:
\begin{align}
\label{eq:alt}
X \geq  b' Y,
\end{align}
where $Y = 1 + X + \sum_{j=1}^{n-1} X_{i_j}$ and $b' = \frac {b}{b + 1}$.

The above model implies that, in the case when there are no interfering transmission (i.e., when $n=1$), a user transmission is recovered only if $ X \geq b$, i.e., the received SNR has to be sufficiently high.
In other words, a user transmission may not be always recovered from a degree one slot, as it is assumed in the simplified SIC scenario, outlined in Section~\ref{sec:CSA}.
Also, it is straightforward to show that, with Rayleigh fading, the probability that a user transmission is successfully recovered from a singleton slot is: 
\begin{align}
\Pr [X \geq b] = e^{-\frac{b}{\bar{\gamma}}}.
\label{eq:deg_one}
\end{align}

\section{Analysis}
\label{sec:analysis}

\begin{figure}[t]
 \begin{center}
  \includegraphics[width=0.6\columnwidth]{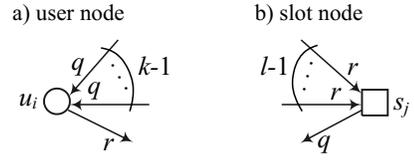}
  \caption{Message updates of the and-or tree evaluation.}
  \label{fig:probs}
 \end{center}
\end{figure}

And-or tree evaluation \cite{LMS1998} is a standard tool used for derivation and assessment of the asymptotic performance of erasure-correcting codes that when decoded by the iterative BP algorithm.
As such, it can be applied for derivation of the asymptotic performance (i.e., when $N \rightarrow \infty$) of coded slotted ALOHA, as presented in \cite{L2011,SPV2012}.
We proceed with a brief overview of the and-or evaluation, to the extent necessary for a seamless incorporation of the capture effect.
For the general introduction on the and-or tree evaluation, we refer the interested reader to \cite{LMS1998,RU2007}.

And-or tree evaluation is concerned with the evaluation of the probabilities that the left-side nodes (i.e, user transmissions) remain unknown (i.e., unrecovered) through the iterations of the SIC algorithm, see Fig.~\ref{fig:graph}.
This is modeled through the exchange of messages flowing between user and slot nodes and carrying the information about the state of the corresponding transmission: not recovered/recovered, which is described with a message value 0/1, respectively.
In each iteration, the probability that a message value is 0/1 is updated according to the following rules.

Consider a user $u_i$, who has transmitted $k$ replicas of the  packet, see Fig.~\ref{fig:probs}a), and assume that the probability that the incoming message value is 0 is $q$, i.e., a replica has not been resolved with probability $q$.
The probability that the value of the outgoing message is $0$ is:
\begin{align}
\label{eq:r}
r = q^{k-1},
\end{align}
i.e., the value of a outgoing message on a edge is 0 only if all incoming messages on the other edges are 0 (the ``or'' update rule).
Averaging \eqref{eq:r} over $k$ yields:
\begin{align}
\label{eq:user}
r_m & =  \sum_k \lambda_k q_{m-1}^{k-1},
\end{align}
where $m$ denotes the iteration, $m \geq 1$,\footnote{It is assumed that $r_0=1$.} and $\lambda_k$ is the probability that a message stems from a node of degree $k$:
\begin{align}
\label{eq:user_edge}
\lambda_k & = \frac{ k \Lambda_k }{\sum_{v} v \Lambda_v},
\end{align}
where $\Lambda_v$ is probability that a user performed $v$ transmissions.
Note that \eqref{eq:user} is the same as in the standard and-or tree evaluation framework; the impact of the capture effect is expressed in the message updates performed in slots.

Consider a slot $s_j$ whose degree is $l$, see Fig.~\ref{fig:probs}b).
The probability that the value of an outgoing message is 1 is:
\begin{align}
\label{eq:liva}
1 - q = \sum_{t=0}^{l-1} \pi_{t}^{l-1} { l - 1 \choose t } ( 1 - r )^{l-1-t}r^{t},
\end{align}
where $\pi_{t}^{l-1}$ expresses probability that a user transmission is recovered in a slot of degree $l$, when $l - 1 - t$ out of $l-1$ interfering transmissions have been canceled due to SIC (i.e., $t$ out of $l-1$ interfering transmissions remain). 
More specifically, $\pi_{t}^{l-1}$ for $1 \leq t \leq l - 1$ represents the contribution of the capture effect that may happen on the yet ``unknown'' messages and which may lead to the user recovery. The combinatorial expression ${ l - 1 \choose t }$ is due to the symmetry of the problem setting: the received SNRs of all interfering transmissions are IID random variables and the occurrence of the ``appropriate'' capture effect on any $t$ out of $l-1$ interfering transmissions is a priori equally likely. 
We note that \eqref{eq:liva} was introduced in \cite{L2011}; also, setting $\pi_{0}^{l-1} = 1$ and $\pi_{t}^{l-1} = 0$ for $1 \leq t \leq l - 1$ yields the standard ``and'' update rule, when it is assumed that there are no noise and no captures. 

Since perfect SIC is assumed, it is easy to show that:
\begin{align}
\label{eq:zero}
\pi_{0}^{l-1} & = e^{-\frac{b}{\bar{\gamma}}}, \\
\pi_{t}^{l-1}  & = \sum_{h=1}^{t+1} C_h^{t+1}, \;1 \leq t \leq l - 1,
\end{align}
where \eqref{eq:zero} stems from \eqref{eq:deg_one}, and where $C_h^{t+1}$ is the probability of the event $\Xi_h^{t+1}$, defined in the following way: at least $h$ captures occurred in the slot of degree $t+1$, among these is the capture related to the user transmission which corresponds to the outgoing message, and this capture occurred as the $h$-th capture.
It is easy to verify that events $\Xi_h^{t+1}$, $1 \leq h \leq t+1 $, are mutually exclusive; we proceed by characterizing the probabilities $C_h^{t+1}$. 

Denote by $X$ the received SNR of the user transmission corresponding to the outgoing message, and by $X_{i_j}$, $1 \leq j \leq t$, the received SNRs of the $t$ interfering users.
Due to the symmetry of the problem setting and the perfect SIC, the probability of $\Xi_h^{t+1}$ is:
\begin{align}
\label{eq:bigC}
C_h^{t+1} & =  \Pr [ \Xi_h^{t+1} ] = \frac{t!}{(t - h + 1)!} \cdot \nonumber \\
& \Pr [ X_{i_1} \geq b' Y_{i_1} \geq  ... \geq X_{i_{h-1}} \geq b' Y_{i_{h-1}} \geq X \geq b' Y_{i_h} ],
\end{align}
where $b'=\frac{b}{b+1}$ and $Y_{i_j} = 1 + X + \sum_{f=j}^{t} X_{i_f} $, $1 \leq j \leq h$, see \eqref{eq:alt}.
In other words, any ordering of received SNRs such that $X$ is the $h$-th largest is a priory likely, which is reflected in \eqref{eq:bigC}.
At this point we note that the computation of \eqref{eq:bigC} in general case is a challenge in its own right, which can be solved using the evaluation method presented in \cite{ZZ2012}, and refer the interested reader to this work for the details.
Also, for $h=1$ it can be shown that in the Rayleigh fading case:
\begin{align}
\label{eq:capture}
C_1^{t+1} = \Pr [ X > b ( 1 + \sum_{j=1}^{t} X_{i_j} )] = \frac{e^{-\frac{b}{\bar{\gamma}}}}{(1+b)^{t} },
\end{align}
where we used the fact that $\sum_{j=1}^{t} X_{i_j} $ is a random variable with gamma distribution $\Gamma(t,\bar{\gamma})$. 

Averaging \eqref{eq:liva} over slot degrees leads to:
\begin{align}
\label{eq:q_update}
q_m = & 1 - \sum_l \omega_l \sum_{t=0}^{l-1} { l - 1 \choose t } \pi_{t}^{l-1} ( 1 - r_m )^{l-1-t}r_m^{t},
\end{align}
where $m$ denotes the iteration, $m \geq 0$, and $\omega_l$ is the probability that a message stems from a slot of degree $l$:
\begin{align}
\label{eq:slot_edge}
\omega_l = \frac{l \Omega_l}{\sum_v v \Omega_v},
\end{align}
where $\Omega_v$ is the probability that slot degree is $v$.

Using \eqref{eq:lambda} and \eqref{eq:omega} specializes \eqref{eq:user} and \eqref{eq:q_update} for the frameless ALOHA:
\begin{align}
\label{eq:frr}
r_m  = & e^{-(1+\epsilon)\beta(1-q_{m-1})}, \\
\label{eq:frq}
q_m  = & 1 - e^{-( \frac{b}{\bar{\gamma}} + \beta r_{m} )} - \nonumber \\ 
- & e^{ - \beta} \sum_l \beta^{l - 1} \sum_{t=1}^{l-1} \pi_{t}^{l-1} \frac { ( 1 - r_m )^{l-1-t}r_m^{t} }{(l-1-t)! t!}.
\end{align} 

Finally, the asymptotic probability of user resolution $P_R$ and the expected throughput $T$ are computed as:
\begin{align}
\label{eq:Pr}
P_R & = 1 - \lim_{m \rightarrow \infty} r_m, \\
\label{eq:Ta}
T & = \frac{P_R}{ 1 + \epsilon}.
\end{align}
We conclude by noting that $P_R$ and $T$ show the \emph{expected} asymptotic performance as functions of the statistical descriptions both of the graph and the capture effect, and as such they are not related to the frameless stopping criterion, as introduced in Section~\ref{sec:model}.

\section{Results}
\label{sec:results}

\subsection{Asymptotic Performance}
\label{sec:asymptotic}

\begin{figure}[t]
 \begin{center}
  \includegraphics[width=\columnwidth]{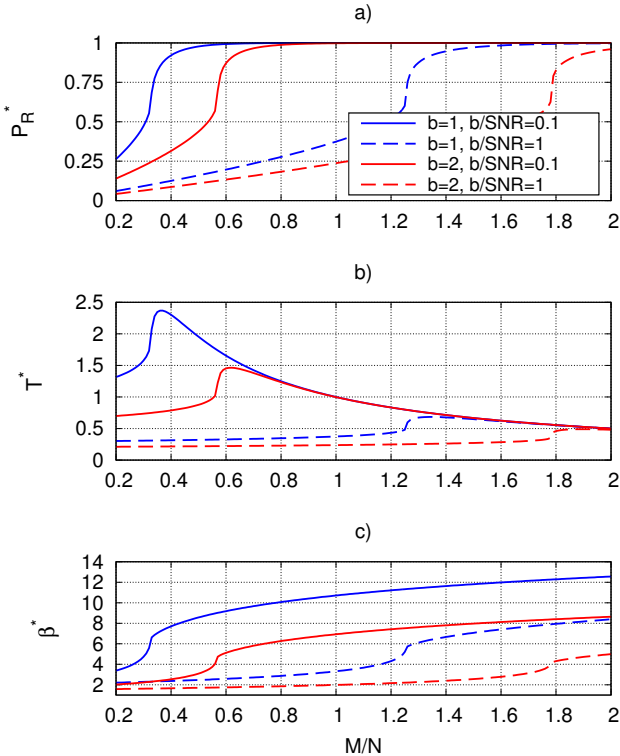}
  \caption{ a) Maximum probability of user resolution $P_R^*$, b) maximum expected throughput $T^*$, and c) the corresponding optimal expected slot degree $\beta^*$, as functions of the ratio of number of slots and number of users $M/N$, for capture threshold $b \in \{1,2\}$ and ratio of the capture threshold to average SNR $b/ \bar{\gamma} \in \{0.1 , 1\}$.}
  \label{fig:and-or}
 \end{center}
\end{figure}

Fig.~\ref{fig:and-or} shows the asymptotic performance obtained by the and-or tree evaluation: a) the maximum probability of user resolution $P_R^*$, b) the corresponding maximum expected throughput $T^*$, and c) the optimum average slot degree\footnote{Which determines the optimum slot access probability, see \eqref{eq:pa}.} $\beta^*$ for which $P_R^*$ and $T^*$ are achieved, as functions of $M/N$.
The results are presented for $b\in \{1,2\}$ and ratio of the capture threshold to the average SNR $b/\bar{\gamma} \in \{0.1,1\}$.
As expected, the increase in $b$ adversely affects the throughput.
Also, for fixed $b$, increase in $b/\bar{\gamma}$ lowers the throughput; this could be expected as well, as lower $b/\bar{\gamma}$ implies: 1) higher probability of recovering a user transmission from a degree one slot, see \eqref{eq:deg_one} and \eqref{eq:zero}, and 2) more chance for the alignment of the received SNRs such that capture occurs in higher-degree slots, cf. (\ref{eq:capture}).
Comparing the results for $\beta^*$ shows that higher throughput is obtained for higher average slot degrees, i.e., the higher slot-access probabilities, see \eqref{eq:pa}, which could be expected as well.
Finally, Fig.~\ref{fig:and-or} shows that the ratio $M/N$ for which the maximum throughput occurs, decreases as this maximum increases.
This is due to the behavior of $P_R^*$, i.e., the sooner $P_R^*$ starts to rise, the higher the $T^*$, see Fig.~\ref{fig:and-or}a) and \eqref{eq:Ta}.
On the other hand, the behavior of $P_R^*$ reflects the fact that, for more pronounced capture effect (i.e., for lower $b$ and lower $b/\bar{\gamma}$) and adequate $\beta$, more users are resolved sooner.

The values of the overall maximum throughput $T^*_{\max}$ and the corresponding $P_R^*$, $\beta^*$ and $M/N$ from Fig.~\ref{fig:and-or} are listed in Table~\ref{tab:asymptotic}, and compared to a scenario where the impacts of both capture effect and noise are neglected \cite{SPV2012}.
Obviously, when the impact of noise is low, i.e., low $b/\bar{\gamma}$, see \eqref{eq:deg_one}, capture effect provides for substantially higher throughputs compared to the scenario without capture effect.
In the case with a considerable impact of the noise, i.e., when $b/\bar{\gamma} = 1$, the probability of recovering transmission from a degree one slot is only $0.37$, see \eqref{eq:deg_one}; this adversely impacts the asymptotically achievable throughput, as shown in Table~\ref{tab:asymptotic}.
We conclude by noting that the optimal $\beta^*$ is substantially higher in scenarios with capture effect, i.e., capture effect favors more collisions per slot.

\renewcommand{\arraystretch}{1.2}
\begin{table}[t]
	\centering
		\begin{tabular}{|c||c|c|c|c||c|}
			\hline
			$b$ & \multicolumn{2}{|c|}{$1$} & \multicolumn{2}{|c||}{$2$} & no capture \\
			\cline{1-5}
			 $b/\bar{\gamma}$ & $0.1$ & $1$ & $0.1$ & $1$ &  effect \\
			\hline
			\hline
			$T^*_{\max}$ & 2.37 & 0.68 & 1.46 & 0.49  & 0.87 \\
			\hline
			$P_{R}^*$ & 0.85 & 0.92 & 0.91 & 0.93 & 0.93 \\
			\hline
 			$\beta^*$ & 7.2 & 6.37 & 5.29 & 4.69 & 3.12 \\
 			\hline
			$M/N$ & 0.36 & 1.34 & 0.62 & 1.89 & 1.07 \\
			\hline
			\end{tabular}
	\caption{Maximum throughput $T^*_{\max}$ and the corresponding $P_R^*$, $\beta^*$ and $M/N$ from Fig.~\ref{fig:and-or}.}
	\label{tab:asymptotic}
\end{table}

\subsection{Non-Asymptotic Performance}
\label{sec:non-asymptotic}

The results in this section are obtained as follows.
For each combination of $N$, $b$ and $b/\bar{\gamma}$, the throughput is maximized over the parameters $\beta$, $S$ and $V$, which are varied with a step $0.01$ through the range of interest.
For each tuple $\beta, S, V$, 10000 simulation runs are made; each run is executed until $F_R \geq V$ or $T_I \geq S$, when $F_R$, $T_I$ and the number of slots $M$ are recorded and then averaged over runs.
$\bar{T}_{\max}$ is the maximum of the average throughput. The optimal values of the parameters that yield $\bar{T}_{\max}$ are denoted as $\beta^*$, $S^*$ and $V^*$. 

Table~\ref{tab:performance} lists the maximum average throughputs $\bar{T}_{\max}$, the corresponding average fraction of resolved users $\bar{F}_R$, the optimal expected slot degrees $\beta^*$, and the normalized average length of the contention period $\bar{M/N}$.
In addition, Table~\ref{tab:thresholds} lists the corresponding optimal values of $V^*$ and $S^*$.

\renewcommand{\arraystretch}{1.2}
\begin{table}[t]
	\centering
		\begin{tabular}{|c|c||c|c|c|c||c|}
			\cline{2-7}
			\multicolumn{1}{c|}{} & $b$ & \multicolumn{2}{|c|}{$1$} & \multicolumn{2}{|c||}{$2$} & no capture effect\\
			\cline{2-6}
			\multicolumn{1}{c|}{} & $b/\bar{\gamma}$ & $0.1$ & $1$ & $0.1$ & $1$ &  and no noise \\
			\cline{2-7}
			\hline
			\multirow{4}{*}{$N=100$} & $\bar{T}_{\max}$ & 1.92 & 0.4 & 1.21 & 0.31 & 0.8 \\
			\cline{2-7}
			& $\bar{F}_R$ & 0.77 & 0.06 & 0.8 & 0.06 & 0.94 \\
			\cline{2-7}
			& $\beta^*$ & 6.14 &  2.23 & 4.53 & 1.55 & 2.89  \\
			\cline{2-7}
			& $\bar{M}/N$ & 0.38 & 0.22  & 0.64 & 0.3 &  1.17 \\
			\hline
			\hline
 			\multirow{4}{*}{$N=1000$} & $\bar{T}_{\max}$ & 2.13 &  0.42 & 1.33 & 0.32  & 0.86 \\
			\cline{2-7}
			& $\bar{F}_R$ & 0.78 & 0.03 & 0.81 & 0.04 & 0.93 \\
			\cline{2-7}
			& $\beta^*$ & 6.91 & 2.38 & 5.1 & 2.15 & 3.04 \\
			\cline{2-7}
			& $\bar{M}/N$ & 0.36 & 0.1  & 0.61 & 0.18 &  1.08 \\			
			\hline
			\end{tabular}
	\caption{Performance of the proposed scheme for number of users $N\in\{100,1000\}$, capture threshold $b \in \{1,2\}$ and ratio of the capture threshold to average SNR $b/ \bar{\gamma} \in \{0.1 , 1\}$.}
	\label{tab:performance}
\end{table}

Obviously, the throughput performance follows the same trends identified by the and-or tree evaluation: it decreases as $b$ increases, for fixed $b$ decreases as $b / \bar{\gamma}$ increases, and, the higher the throughput, the higher $\beta^*$ required to achieve it.
Further, the throughput increases as $N$ increases; the same behavior was observed in \cite{SPV2012,SP2013}, and it is due to the well-known fact that rateless codes perform better as the length of the information sequence increases.

However, when compared to the results presented in Table~\ref{tab:asymptotic}, it is obvious that the non-asymptotic throughput values are lower.
This can be explained by the fact that the and-or tree evaluation assumes that the underlying graph on which SIC is performed is a tree.
On the other hand, in the non-asymptotic case, the actual graph contains loops which cause interdependencies among slots' contents and which may prevent recovery of the related user transmissions; these loops are shorter when the number of users is lower.
The interdependencies among slots with respect to SIC are further amplified by the fact that in the assumed system model the received SNR of a particular user has the same value in all slots in which user transmitted; this detail is also not considered in the and-or tree evaluation.
The discrepancy between the asymptotic and non-asymptotic performance is particularly pronounced for low $b/\bar{\gamma}$, when the ``potential'' both of the degree one slots and the capture effect is low.
Also, it could be shown that for low $b/\bar{\gamma}$, in most of the cases the length of the contention period $M$ is only a few slots and the contention is terminated when $T_I \geq S^*$; this can be also observed by inspecting the values of $\bar{F}_R$, $\bar{M}/N$, $V^*$, and $S^*$ in Tables~\ref{tab:performance} and \ref{tab:thresholds}.
In other words, the instances in which few user transmissions have been resolved at the beginning of the contention period are immediately terminated, as the attained throughput is likely only to decrease.

\section{Conclusions}
\label{sec:conclusions}

In this paper we analyzed the impact of the capture effect on the SIC-enabled slotted ALOHA framework, with an emphasis on the case of frameless ALOHA in Rayleigh fading scenario.
We have shown that a pronounced capture effect boosts the potential of the SIC, by favoring a higher number of colliding user in a slot, and thereby leading to considerably higher throughput. 
These results motivate further investigation of coded random access, but with further refinement of the models at the physical layer, such as actual modulation and coding.

\section*{Acknowledgment}

The research presented in this paper was partly supported by the Danish Council for Independent Research (Det Frie Forskningsr{\aa}d) within the Sapere Aude Research Leader program, Grant No. 11-105159 ``Dependable Wireless Bits for Machine-to-Machine (M2M) Communications''.

\bibliographystyle{IEEETran}


\begin{table}[t]
	\centering
		\begin{tabular}{|c|c||c|c|c|c||c|}
			\cline{2-7}
			\multicolumn{1}{c|}{} & $b$ & \multicolumn{2}{|c|}{$1$} & \multicolumn{2}{|c||}{$2$} & no capture effect\\
			\cline{2-6}
			\multicolumn{1}{c|}{} & $b/\bar{\gamma}$ & $0.1$ & $1$ & $0.1$ & $1$ & and no noise \\
			\cline{2-7}
			\hline
			\multirow{2}{*}{$N=100$} & $V^*$ & 0.7 & 0.14 & 0.74 & 0.14 & 0.88 \\
			\cline{2-7}
			& $S^*$ & 2.02 & 0.34 & 1.3 & 0.25 & 0.81 \\
			\hline
			\hline
 			\multirow{2}{*}{$N=1000$} & $V^*$ & 0.74 &  0.1 & 0.78 & 0.12 & 0.89 \\
			\cline{2-7}
			& $S^*$ & 2.19 & 0.34 & 1.35 & 0.25 & 0.87 \\
			\hline
			\end{tabular}
	\caption{Optimal $V^*$ and $S^*$ for which the performance in Table~\ref{tab:performance} is obtained.}
	\label{tab:thresholds}
\end{table}

\end{document}